\documentclass[twocolumn,showpacs,preprintnumbers,amsmath,amssymb]{revtex4}
\usepackage{graphicx}
\begin{document}

%\preprint{}

\title{Distinct Pairing Symmetries in $Nd_{1.85}Ce_{0.15}CuO_{4-y}$ and $La_{1.89}Sr_{0.11}CuO_{4}$
Single Crystals: Evidence from Comparative Tunnelling Measurements}

\author{L. Shan$^1$, Y. Huang$^1$, H. Gao$^1$, Y. Wang$^1$, S. L. Li$^2$, P. C. Dai$^{2,3}$, F. Zhou$^1$, J. W.
Xiong$^1$, W. X. Ti$^1$, }
\author{H.H. Wen$^1$}

\affiliation{$^1$National Laboratory for Superconductivity, Institute of Physics, Chinese
Academy of Sciences, P.O. Box 603, Beijing 100080, China}

\affiliation{$^2$Department of Physics and Astronomy, The University of Tennessee,
Knoxville, Tennessee 37996-1200, USA}

\affiliation{$^3$Center for Neutron Scattering, Oak Ridge National Laboratory, Oak Ridge, Tennessee
37831-6393, USA }

\date{\today}

\begin{abstract}
We used point-contact tunnelling spectroscopy to study the superconducting pairing
symmetry of electron-doped $Nd_{1.85}Ce_{0.15}CuO_{4-y}$ (NCCO) and hole-doped
$La_{1.89}Sr_{0.11}CuO_{4}$ (LSCO). Nearly identical spectra without zero bias
conductance peak (ZBCP) were obtained on the (110) and (100) oriented surfaces (the
so-called nodal and anti-nodal directions) of NCCO. In contrast, LSCO showed a remarkable
ZBCP in the nodal direction as expected from a $d$-wave superconductor. Detailed analysis
reveals an $s$-wave component in the pairing symmetry of the NCCO sample with
$\Delta/k_BT_c=1.66$, a value remarkable close to that of a weakly coupled BCS
superconductor. We argue that this s-wave component is formed at the Fermi surface
pockets centered at ($\pm\pi$,0) and (0,$\pm\pi$) although a $d$-wave component may also
exist.

\end{abstract}


\maketitle

\section{Introduction}

For hole-doped cuprate superconductors, there is a well-known electronic phase diagram
characterized by a dome-like superconducting region when $p$ is above a certain threshold
\cite{OrensteinJ2000}. It was found that \cite{TokuraY1989} instead of doping holes into
the Cu-O plane, one can also achieve superconductivity by doping electrons into the Cu-O
plane in systems like Ln$_{2-x}$Ce$_x$CuO$_4$ with Ln = Nd, Pr, La, etc. One thus
naturally questions whether the superconducting mechanism is the same between the hole-
and electron- doped cuprates. Among many superconducting properties, the symmetry of the
order parameter is an important one. While the symmetry of the superconducting order
parameter is believed to be $d$-wave for the hole-doped region \cite{TsueiCC2000}, the
situation for electron-doped materials is highly controversial. For example,
Angle-resolved photoemission spectroscopy (ARPES) \cite{ArmitageNP2001}, specific heat
\cite{BalciH2002}, phase-sensitive scanning SQUID \cite{TsueiCC2000PRL}, bi-crystal
grain-boundary Josephson junction\cite{Chesca} and some penetration depth measurements
\cite{Penetrationdepth01,SnezhkoA2004} indicate a $d$-wave symmetry. In addition, Raman
scattering \cite{BlumbergG2002} and recent ARPES \cite{MatsuiH2004} experiments show a
nonmonotonic $d$-wave order parameter. However, this has been contrasted by tunnelling
\cite{ChenCT2002,KashiwayaS1998} and some other specific heat \cite{LiuZY2004} and
penetration depth measurements \cite{AlffL1999,SkintaJA2002,KimMS2003}. In particular,
there may be a crossover from $d$-wave to $s$-wave symmetries by changing the doped
electron concentration \cite{SkintaJA2002b,BiswasA2002} or decreasing temperature
\cite{BalciH2004}. Although such a crossover may explain the conflicting experimental
results on the pairing symmetry in the electron-doped cuprates, the characteristics of
such possible $s$-wave pairing symmetry has yet to be established.

In this paper, we report directional tunnelling measurements on single crystals of
optimally electron-doped cuprate $Nd_{1.85}Ce_{0.15}CuO_{4-y}$ (NCCO). By injecting
current along either the Cu-Cu bond (110) or Cu-O bond (100) direction, we obtain nearly
identical tunnelling spectra indicating an $s$-wave component of the pairing symmetry in
this material. For comparison, similar measurements were carried out on
 underdoped $p$-type single crystals of
$La_{1.89}Sr_{0.11}CuO_{4}$ (LSCO), and we observe clear zero bias conductance peaks
(ZBCP) in tunneling spectra along the (110) direction as expected from the
$d_{x^2-y^2}$-wave symmetry. Our results thus indicate that the optimally doped NCCO has
at least an unavoidable $s$-wave component, which is in contrast with the case in
hole-doped LSCO where a pure $d$-wave has been established.

\section{Experimental details}

\begin{figure}[bottom]
\includegraphics[scale=1.05]{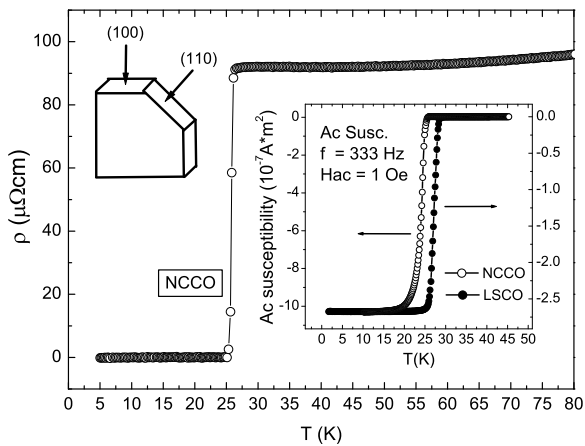}
\caption{\label{fig:fig1} The temperature dependence of resistivity for the sample NCCO.
The insets show the AC susceptibility of NCCO and LSCO. The schematic diagram of the
point contact configuration are presented in the upper left of the main panel.}
\end{figure}

We grew the NCCO and LSCO single crystals using the travelling-solvent floating-zone
technique \cite{OnoseY2001,ZhouF2003}. As shown in Fig.~\ref{fig:fig1}, the resistive
curve measured on the NCCO sample indicates a zero-resistance temperature at 25.1K, AC
susceptibility shows the onset of bulk superconductivity at $T_c\approx 25.6$K. The LSCO
sample has a $T_c\approx 28$K characterized by AC susceptibility measurement. The single
crystals used in the tunnelling measurements were cut into rectangular flakes of size
$2.5\times 2.5\times 1$mm$^3$ for NCCO and $1.5\times 1.5\times 0.8$mm$^3$ for LSCO with
the long axis along $a$ (100) or $b$ (010) and short axis along $c$ (001), one corner of
the crystal was then cut off to expose (110)-oriented crystal planes. The directional
point-contact tunneling measurements were carried out by pointing a Pt/Ir alloy tip or Au
tip towards the specified directions as shown in the left inset of Fig.~\ref{fig:fig1}.
The tip's preparation and the equipment details are described in Ref.\cite{ShanL2003}. In
order to reduce quasiparticle scattering in the barrier layer and hence obtain high
quality data, the nonaqueous chemical etch was used to attenuate the insulating layer on
the sample surface immediately before mounting the sample on the point contact device.
Since it takes about 20-40 minutes to insert the sample mounted device into the Helium
gas environment in the dewar, the sample surfaces were exposed to air during this period.
For chemical etching, we use $10\%$ HCl in absolute ethanol for several seconds to
several tens of seconds for NCCO and $1\%$ Br$_2$ in absolute ethanol for several minutes
to ten minutes in the case of LSCO. Typical four-terminal and lock-in technique were used
to measure the $I\sim V$ curves and the differential resistance $dV/dI$ vs $V$ of the
point contacts.

\section{Results and discussions}
\subsection{Tunneling spectra of Nd$_{1.85}$Ce$_{0.15}$CuO$_{4-y}$ (NCCO)}

\begin{figure}[]
\includegraphics[scale=0.85]{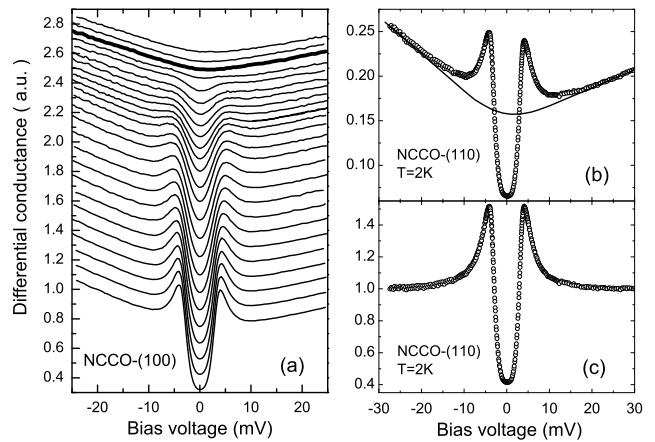}
\caption{\label{fig:fig2} Raw data of the directional spectral measurements. (a) The
temperature dependent tunneling spectra measured along (100) direction. The curves have
been shifted for clarity. The temperature increases from the bottom up with the steps of
1K (from 2K to 22K) and 2K (from 22K to 30K). The thick solid line denotes the data at
26K which is around $T_c$. (b) An illustration of constructing normal conductance
background according to the functional form of the spectra above $T_c$. (c) Normalized
spectrum for (110) direction at $T=2$K. }
\end{figure}

\begin{figure}[]
\includegraphics[scale=1.2]{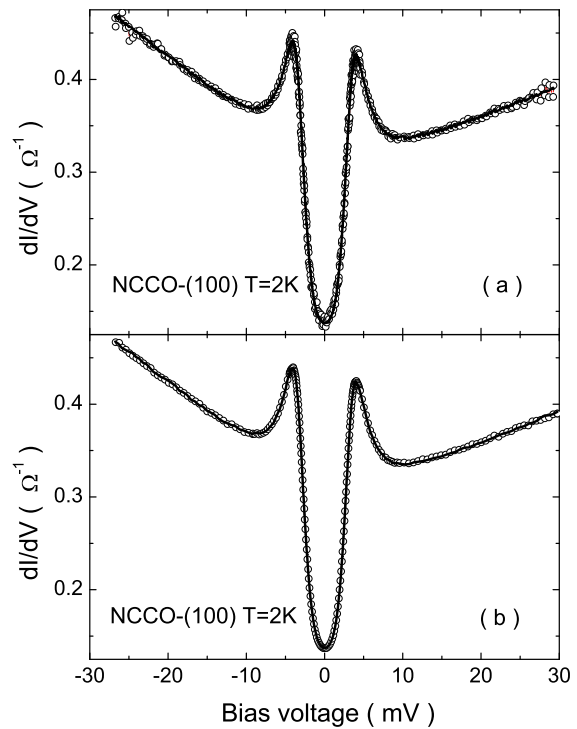}
\caption{\label{fig:fig3} Comparison between the $\sigma\sim V$ curves (a) obtained from
lock-in technique (solid line) and DC $I\sim V$ measurements (open circles); (b) measured
with positive (solid line) and negative (open circles) bias scanning.}
\end{figure}
Fig.~\ref{fig:fig2} shows the raw data of the conductance ($\sigma$) of Au/NCCO point
contact for various temperatures from 2K to 30K. In order to show the electronic
stability of the experiments, we simultaneously present in Fig.~\ref{fig:fig3}(a) the
$\sigma\sim V$ curves obtained from the lock-in technique and DC $I\sim V$ measurement,
respectively. It is obvious that these two curves merge into each other although the
latter one has a higher noise. We also compared the $\sigma\sim V$ curves recorded by
both positively and negatively scanning the bias voltage. As shown in
Fig.~\ref{fig:fig3}(b), the complete overlap of the data with different scanning
directions indicates the good thermal stability during measurements.

\begin{figure}[]
\includegraphics[scale=1.2]{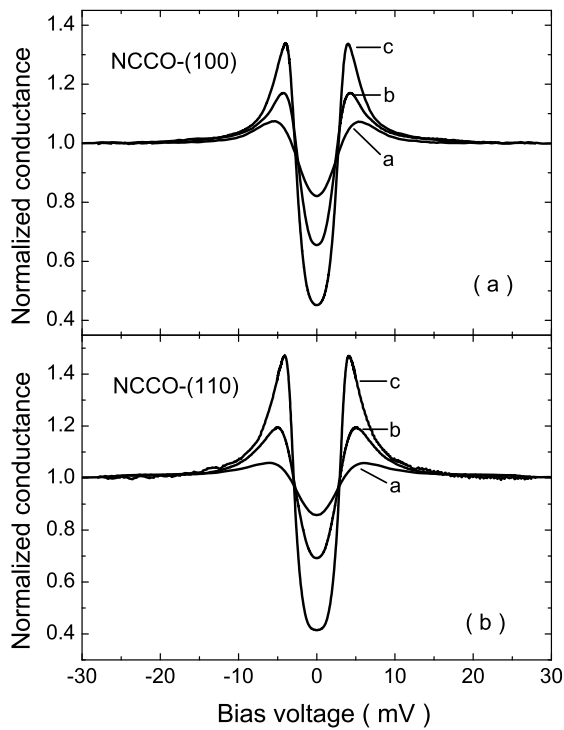}
\caption{\label{fig:fig4} The normalized $\sigma\sim V$ curves obtained by different
surface treatments and using different tips. Note that the spectra measured on the
post-etched surfaces are much sharper than that of pre-etched ones, accompanied by a
remarkable decrease in the junction resistance from several hundreds of Ohms to below 10
$\Omega$ as shown in Table~\ref{tab:table1}. }
\end{figure}

\begin{table}
\caption{\label{tab:table1} List of the experimental conditions for the different spectra
presented in Fig.~\ref{fig:fig4}.}
\begin{ruledtabular}
\begin{tabular}{lccc}
Label & Chemical Etch & Tip & $R_{20mV}$ ($\Omega$) \\
\hline
(110)-a       & None (As cut)                               & Pt/Ir Alloy &197.0  \\
(110)-b       & By $10\%$ HCl (20s)  & Pt/Ir Alloy &75.7  \\
(110)-c       & By $10\%$ HCl (60s)  & Au          &5.3  \\
\hline
(100)-a       & None (As cut)                               & Pt/Ir Alloy &170.0  \\
(100)-b       & By $10\%$ HCl (20s)  & Pt/Ir Alloy &19.9  \\
(100)-c       & By $10\%$ HCl (60s)  & Au          &2.9  \\
\end{tabular}
\end{ruledtabular}
\end{table}
It is well known that the high $T_c$ superconducting cuprate compounds react readily with
atmospheric H$_2$O and CO$_2$ to form insulating hydroxides and carbonates on the surface
and at grain boundaries \cite{VasquezRP1988}. While such poor conductive layer may be a
natural barrier needed in tunneling experiments, they may introduce large scattering
factor (just like the situation of fabricated planar junction) and thus result to less
sharpening spectra than those measured by the vacuum barrier based STM \cite{LiSX2000}.
In order to reduce the scattering factor and obtain sharp spectra, we used nonaqueous
chemical etch to treat the sample surfaces \cite{VasquezRP1988,VasquezRP1994}. In
Fig.~\ref{fig:fig4}, we present the measured spectra on the surfaces as cut, after the
first etch and after the second etch, all the spectra have been normalized for
comparison. The detailed experimental conditions are listed in Table~\ref{tab:table1}.
The experimental results indicate that the chemical etch can effectively reduce the
thickness of the barrier layer on the surface of NCCO single crystal and hence decrease
the scattering factor.

\begin{figure}[top]
\includegraphics[scale=1.2]{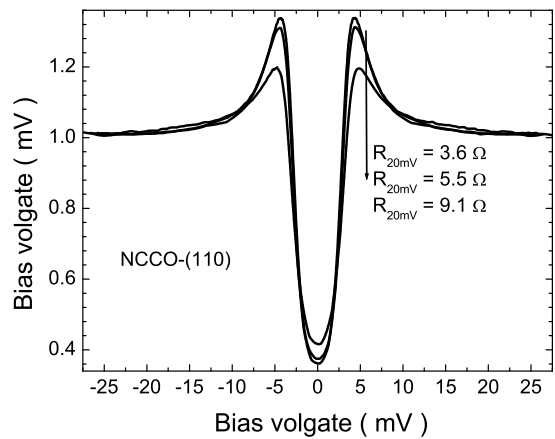}
\caption{\label{fig:fig5} The normalized $\sigma\sim V$ curves measured at a fixed
position on the sample surface with various junction resistance.}
\end{figure}
We have also studied the dependence of the measured spectrum on the junction resistance
which can be controlled by adjusting the tip-to-sample distance with a differential
screw. Fig.~\ref{fig:fig5} shows the spectra measured at a fixed position on the sample
surface as a function of junction resistances, in these measurements the sample surfaces
had been chemically etched. When the junction resistance decreases from several tens of
Ohms to several Ohms, the spectral shape becomes sharper (refer to
Fig.~\ref{fig:fig8}(a)). For small resistance, the spectral shape changes hardly with the
resistive values and no Andreev reflection-like spectrum occurs \cite{BlonderGE1982}
(Fig.~\ref{fig:fig5}). This indicates that although the surface contaminant phase has
been reduced, it still exists as a solid barrier layer due to the exposure to air. After
the metal tip reaches the sample surface, the barrier layer is abraded at first and its
thickness decreases with the increasing pressure. Consequently, the measured spectrum
becomes sharper due to the weakening of the inelastic scattering of the injecting
quasiparticles near the normal-metal/superconductor (N/S) micro-constriction. Further
pressure of the tip on the sample surface may simply flatten the point, giving a larger
contact area over the same minimal thickness of a tenacious barrier layer which can not
be entirely flaked off \cite{BlonderGE1983}. In this case, the junction resistance
becomes still smaller while with no obvious change on the measured spectrum until the
junction is damaged eventually. Therefore, the none-zero zero bias conductance should be
ascribed to the stronger scattering effect on injecting quasiparticles and lower height
of the surface barrier than that of the vacuum barrier. Nonetheless, such surface
contaminants resulting from air exposure form orders of magnitude slower on NCCO than on
many hole-doped cuprates which results from the absence of reactive alkaline earth
elements in NCCO \cite{VasquezRP1994}. This is also the reason why the measured spectra
of LSCO are poorer than that of NCCO using the same method, as shown in the section III-C
in this paper. In the following, we will focus our attention on the spectra with the
resistance from 1 to 15 $\Omega$, namely, in the regime with good spectral stability and
weaker scattering effect.

\begin{figure}[]
\includegraphics[scale=1.2]{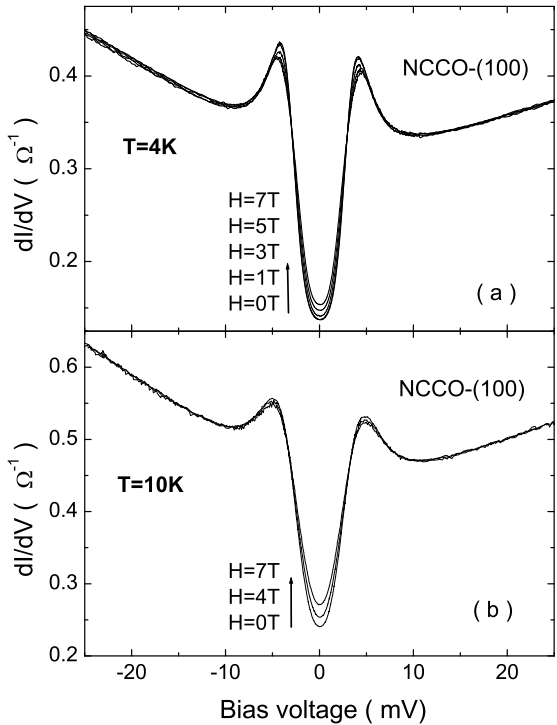}
\caption{\label{fig:fig6} The magnetic field dependence of the tunneling spectra measured
at (a) 4K and (b) 10K. The very small modification of the spectral shape in field up to
7T indicates that we indeed measured the bulk superconductivity of NCCO.}
\end{figure}
In MaglabExa-12, the magnetic field can be applied along the tip's direction, namely,
parallel to the $ab$-plane. The field dependence of the tunneling spectra were measured
at various temperatures from 4K to 16K. The changing tendency of the spectral shape with
increasing field is identical at all measured temperatures. The typical results of 4K and
10K in Fig.~\ref{fig:fig6} suggest that the field induced smearing of the spectrum is
very slight, which is consistent with the much higher $H_{c2}$ than 7T for $H\parallel
ab-plane$.

To sum up the main characteristics of the measured spectra, we note that all the spectra
are near the tunneling limit with two clear coherence peaks at symmetric positions of the
bias voltage (much sharper than that of the previous works studied by both STM
\cite{KashiwayaS1998,KashiwayaS2003} and point contact \cite{MourachkineA2000}) with no
Andreev reflection-like spectrum. Second, the spectral shape is nearly identical for the
(110) and (100) directions. Thirdly, there was no evidence for zero-bias conductance peak
(ZBCP) expected in the (110) (or the so-called nodal direction) of the $d$-wave
superconductors \cite{HuCR1994,TanakaY1995,KashiwayaS2000}. Here we emphasize that the
results are independent of the positions on the sample surface.

\begin{figure}[]
\includegraphics[scale=1.2]{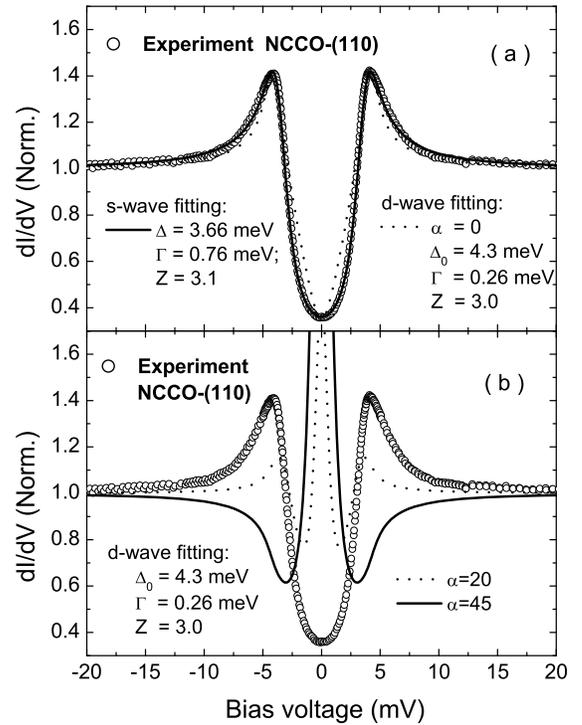}
\caption{\label{fig:fig7} The comparison of fitting the normalized conductance measured
along (110)-direction at 2K to the $s$-wave model and $d_{x^2-y^2}$-wave model. For,
$d$-wave fitting, $\alpha$ denotes the angle between the normal direction of the sample
surface and the crystallographic axis $a$. Note that the $s$-wave fitting is much better
than the $d$-wave fitting. }
\end{figure}

\subsection{Theoretical model - $s$-wave BTK theory}

As discussed above, if the superconducting NCCO has $d$-wave pairing symmetry, a ZBCP
should be observed in the (110)-oriented tunneling spectrum. In other words, our
experimental results suggest that the optimally doped NCCO has a $s$-wave symmetry. In
order to explain this viewpoint, we present in Fig.~\ref{fig:fig7} the best fitting of
the (110)-oriented tunneling spectrum with the formulas corresponding $s$-wave and
$d$-wave models respectively. For the simulations, the extended BTK model
\cite{BlonderGE1982,TanakaY1995} was accepted by selecting a constant gap value for
$s$-wave symmetry and the anisotropic gap of $\Delta(\theta)=\Delta_0cos(2\theta)$ for
$d_{x^2-y^2}$ symmetry, where $\theta$ is the polar angle measured from the
crystallographic axis $a$. In this model, two parameters are introduced to describe the
necessary physical quantities, i.e., the effective potential barrier ($Z$) and the
superconducting energy gap ($\Delta$). As an extension, the quasiparticle energy $E$ is
replaced by $E-i\Gamma$, where $\Gamma$ is the broadening parameter characterizing the
finite lifetime of the quasiparticles due to inelastic scattering near the N/S
micro-constriction \cite{DynesRC1984,PlecenA1994}. In a real N-I-S junction
configuration, total tunneling conductance spectrum includes the integration over the
solid angle. In the case of two dimension, it reduces to the integration over the
injection angle from $-\pi/2$ to $\pi/2$, as done in this work. Fig.~\ref{fig:fig7}(b)
clearly shows that the $d$-wave theoretical simulation along the (110) direction deviates
from the experimental data. When the normal direction of the sample surface departs a
small angle from the crystallographic axis $a$, a ZBCP will appear, accompanied by the
remarkable depression of the coherence peaks. As shown in Fig.~\ref{fig:fig7}(a), even
the $d$-wave simulation absolutely along (100) direction can not fit the experimental
data in the range below superconducting gap. However, the calculated curves in terms of
the $s$-wave theory are in good agreement with the experimental results both in (100) and
(110) directions.

\begin{figure}[]
\includegraphics[scale=1.2]{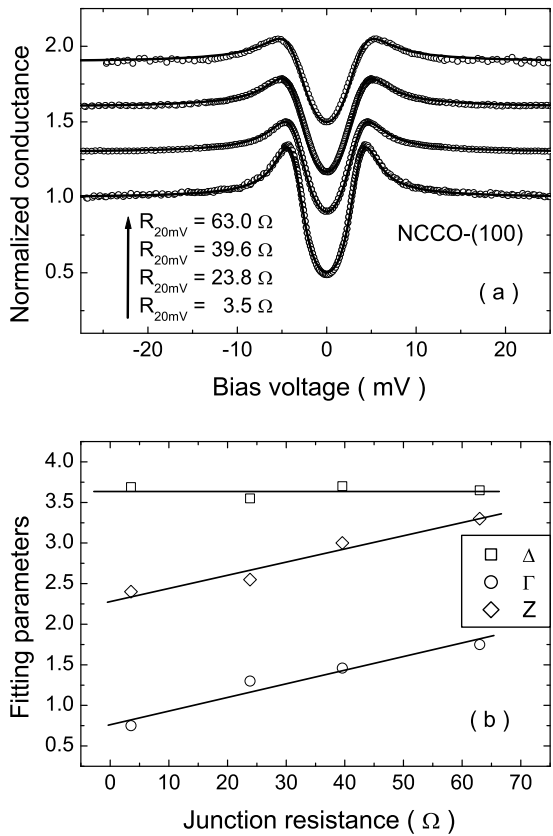}
\caption{\label{fig:fig8} (a) The normalized spectra measured at a fixed position on the
sample surface with various junction resistance in a large resistive range. The open
circles denote the experimental data and the solid lines indicate the fitting to the
$s$-wave extended BTK model. (b) The fitting parameters for the spectra presented in (a).
}
\end{figure}

It should be pointed out that the values of $Z$ and $\Gamma$ are related to the selected
form of the background (such as the slope coefficient of the high bias straight-line
segments) in a certain extent, which may be the main artificial uncertainty in the data
analysis. So we did not focus our attention on the absolute values of these parameters or
the quantitative comparison between the different spots. However, it is also found that
the energy gap $\Delta$ is insensitive to the selection of different background in the
fitting process. Moreover, when we chose a general procedure to construct the background
for all spectra measured on a fixed spot on the sample surface, the variation of the
fitting parameters of $Z$ and $\Gamma$ should be physically meaningful. As an example, we
presented in Fig.~\ref{fig:fig8}(a) the normalized spectra measured at a fixed spot with
different junction resistance, which have been theoretically fitted to the $s$-wave BTK
model and the fitting parameters are shown in Fig.~\ref{fig:fig8}(b). It is noted that
the derived energy gap is nearly constant for all spectra, while the barrier height $Z$
and broadening parameter $\Gamma$ continuously decreases with the decreasing junction
resistance. This is physically reasonable and in good agreement with the foregoing
discussions, namely, the abrasion of the contaminant layer will depress the barrier
height accompanied by the weakening of the quasiparticle scattering near the N/S
micro-constriction.

\begin{figure}[]
\includegraphics[scale=0.85]{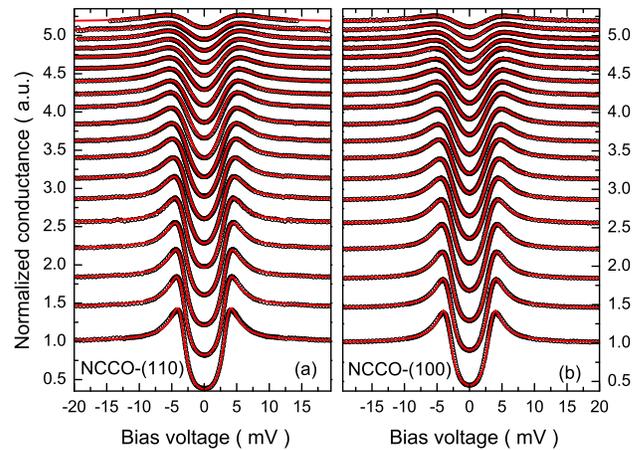}
\caption{\label{fig:fig9} The temperature dependent normalized conductance and the
$s$-wave BTK simulations for (a) (110) direction and (b) (100) direction. All the curves
have been shifted upwards except the lowest one for clarity. The temperature increases
from 2K to 20K with a interval of 1K (from the bottom up). The theoretical simulations
are denoted by red solid lines. }
\end{figure}

\begin{figure}[]
\includegraphics[scale=1.03]{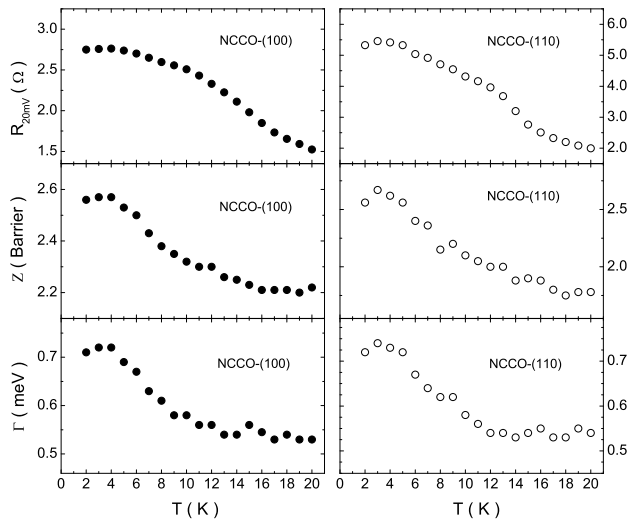}
\caption{\label{fig:fig10} The junction resistance measured at 20mV, the best fitting
parameters of barrier height $Z$ and broadening parameter $\Gamma$ corresponding to the
simulations shown in Fig.~\ref{fig:fig9}. The junction resistances $R_{20mV}$ of
different spots on the sample surface possess the similar decay law with increasing
temperature, indicating that the slight thermal expansion of the point-contact device.
Correspondingly, the values of $Z$ and $\Gamma$ decrease with increasing temperature at
first, then tend to be constant when the junction resistance becomes small enough because
the minimal thickness of the barrier layer is achieved and the point flattening occurs.}
\end{figure}

\begin{figure}[]
\includegraphics[scale=1.2]{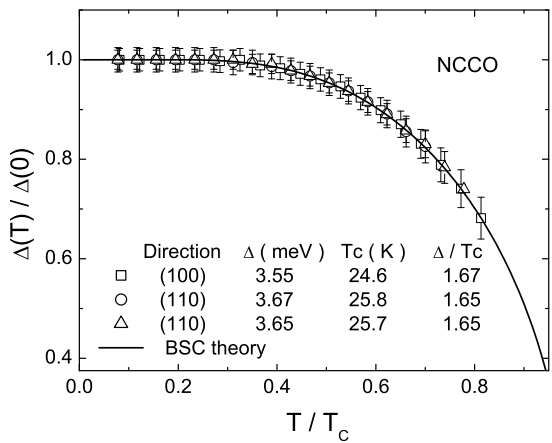}
\caption{\label{fig:fig11} The determined $\Delta(T)$ relations by fitting the
temperature dependent tunneling spectra to the $s$-wave BTK theory. The solid line
indicates the BCS prediction.}
\end{figure}

The good spatial repeatability of the tunneling spectra allowed further investigations on
their temperature dependence. The spectral shape change continuously until above 24K
($T_c\sim26K$), thus suggesting that the tunneling spectra reflect the bulk
superconductivity of the NCCO sample (Fig.~\ref{fig:fig2}(a)). The temperature dependent
normalized spectra for both (110) and (100) directions are presented in
Fig.~\ref{fig:fig9}, accompanied by the $s$-wave BTK fitting curves denoted by the red
solid lines. Because the spectra were affected by the critical current effect at higher
temperature near $T_c$ \cite{SheetG2004}, it is difficult to determine the background for
these temperatures. Therefore, these results have not been included in our theoretical
analysis. All the parameters of the simulations shown in Fig.~\ref{fig:fig9} are
presented in Fig.~\ref{fig:fig10} (the barrier height $Z$ and the broadening parameter
$\Gamma$) and Fig.~\ref{fig:fig11} (superconducting energy gap $\Delta$). The junction
resistances measured at 20mV ($R_{20mV}$) are also given in Fig.~\ref{fig:fig10}.

\begin{figure}[]
\includegraphics[scale=1.2]{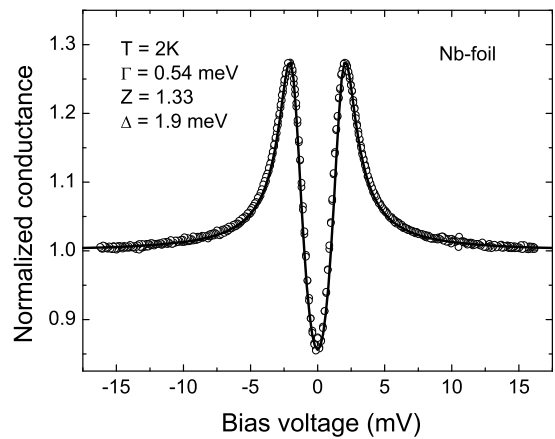}
\caption{\label{fig:fig12} The point contact spectrum measured at 2K on the Nb foil
exposed in air for long time. The solid line denotes the $s$-wave BTK fitting. The
obtained value of $A=\Gamma/\Delta(0)=0.28$ is much larger than that obtained from
Au/NCCO tunneling in this work.}
\end{figure}

The magnitude of $A=\Gamma/\Delta(0)<20\%$ presented is larger than that of the clean
point-contact junction between normal metal and conventional superconductors which in
general is smaller than $10\%$. However, considering the much shorter coherence length
and faster exterior degradation of cuprates than that of the conventional
superconductors, one can easily understand the larger $A$. Actually, a large $A$
($>20\%$) is often obtained on the oxidized surface of Nb foil, a typical conventional
superconductor, as shown in Fig.~\ref{fig:fig12}. The lowest value of $A=17\%$ achieved
in this work is much smaller than the previous report of $A\approx 1$ by STM measurements
on NCCO single crystal \cite{KashiwayaS1998}.

As shown in Fig.~\ref{fig:fig11}, the normalized gap function $\Delta (T)$ determined by
the fitting procedure can be well described by BCS theory, yielding a gap ratio
$\Delta(0)/k_BT_c\approx 1.66$, which is very close to the theoretical value of $1.76$.
Such consistency between the tunneling data of cuprates and the theoretical model over a
wide temperature range (between 0K and $T_c$) is surprising. In order to determine the
repeatability of such temperature dependent measurements, we plot in Fig.~\ref{fig:fig11}
two $\Delta(T)$ relations obtained at two different positions for the (110) direction.
All the $\Delta(T)$ functions obtained along both (110) and (100) directions and from
different positions follow the BCS theory in a normalized scale. The $T_c$ values derived
from this figure is between 24.5K$\sim$26K, near the bulk transition temperature
$T_c\approx 25.6K$.

\begin{figure}[]
\includegraphics[scale=1.2]{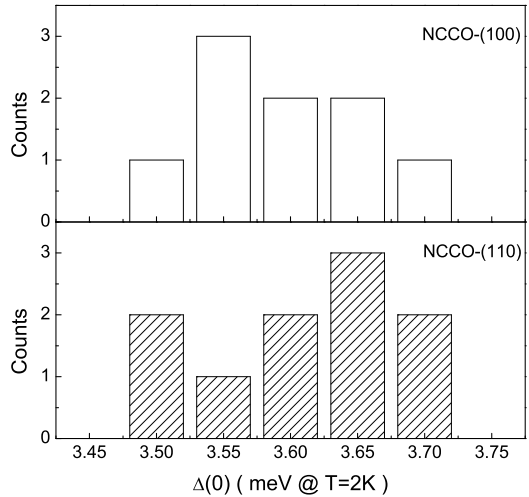}
\caption{\label{fig:fig13} Statistical chart of the $\Delta(0)$-values determined by the
measurements at many different positions on the sample surfaces. The $y$-axis denotes the
occurring times for a given gap-value specified by the $x$-axis. }
\end{figure}

The distribution of the superconducting energy gap $\Delta(0)$ derived from the spectral
measurements at many different positions are shown in Fig.~\ref{fig:fig13}. There is no
obvious difference between the data obtained from (110) and (100) directions. The value
of $\Delta(0)$ varies in a narrow range from 3.50 to 3.70, indicating the good
homogeneity of the superconductivity in the investigated regions.

\begin{figure}[]
\includegraphics[scale=1.2]{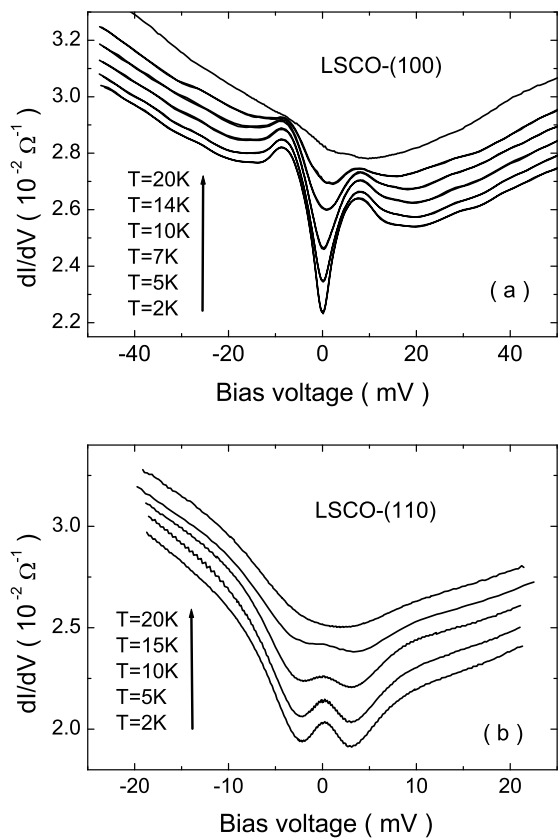}
\caption{\label{fig:fig14} Temperature dependent tunneling spectra of
La$_{1.89}$Sr$_{0.11}$CuO$_{4}$ single crystal measured along (a) (100) direction and (b)
(110) direction. The distinct spectral characteristics of these two directions are
obviously different from the case of NCCO.}
\end{figure}

\subsection{Tunneling spectra of La$_{1.89}$Sr$_{0.11}$CuO$_{4}$ (LSCO)}

As a comparison, the directional tunneling spectra along both (100) and (110) directions
were also studied on the LSCO single crystal, which has a typical $d_{x^2-y^2}$-wave
pairing symmetry (as checked recently by specific heat \cite{WenHH2004}). The measured
spectra at various temperatures are presented in Fig.~\ref{fig:fig14}. For the (100)
direction at low temperatures, two clear coherence peaks appear in the spectra and no
ZBCP can be observed. While for the (110)-oriented spectra, a prominent ZBCP was
observed, accompanied by the disappearance of the coherence peaks. With the increasing
temperature, all these characteristics become weaker and weaker and eventually vanish
around $T_c$. As mentioned above, photoemission experiments have proved that surface
degradation resulting from air exposure form orders of magnitude faster on LSCO than
NCCO, due to the existence of reactive alkaline earth elements in LSCO
\cite{VasquezRP1994}. Therefore, the surface of LSCO degraded quickly before being put
into the helium gas environment, which directly affected the subsequent spectral
measurements. That is, the injecting quasiparticles felt strong scattering effect from
the surface barrier layer (resulting in a very large spectral broadening factor), i.e.,
the measured spectra were badly smeared. As shown in Fig.~\ref{fig:fig14}, the
superconducting characteristics nearly completely disappear above 20K which is much lower
than the bulk critical temperature $T_c\approx 28K$.

\begin{figure}[]
\includegraphics[scale=1.2]{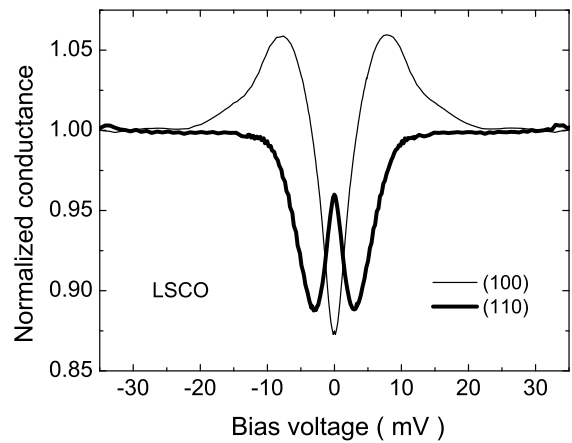}
\caption{\label{fig:fig15} Normalized tunneling spectra of $T=2$K. Thick and thin solid
lines indicate (110) and (100) directions, respectively. The remarkable difference
between the cases along the two axis is in good agreement with the $d$-wave theory.}
\end{figure}

In order to further demonstrate the spectral differences along the two directions, the
spectra of $T=2$K have been normalized using the higher bias backgrounds, as shown in
Fig.~\ref{fig:fig15}. Although the strong scattering effect has severely smeared the
spectra and made the quantitative analysis difficult, the coherence peaks around gap
energy are depressed and a ZBCP appears when the tunneling direction changes from (100)
to (110) as expected from the $d$-wave theory. These spectra are also consistent with the
results obtained on the La$_{2-x}$Sr$_{x}$CuO$_{4}$/Ag junctions fabricated using a
ramp-edge technique \cite{MiyakeT2003}.

\subsection{ Theoretical model - Anisotropic $s$-wave and two-band model}

\begin{figure}[]
\includegraphics[scale=1.2]{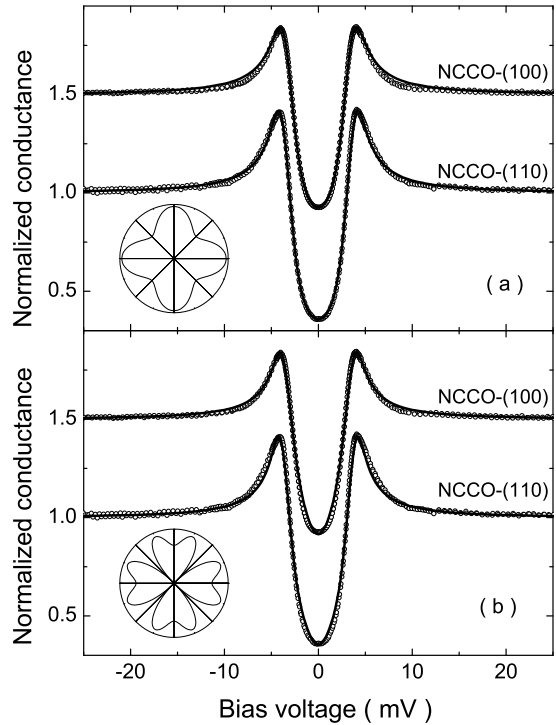}
\caption{\label{fig:fig16} Fitting the directional tunneling spectra to the anisotropic
$s$-wave model with gap functions of (a) $\Delta(\theta)=\Delta_0+\Delta_1cos(4\theta)$
and (b) $\Delta(\theta)=\Delta_0\cdot|cos(2\theta)-0.3cos(6\theta)|$. The insets are the
schematic drawing of the angle dependent energy gaps. The best fitting to the first type
of anisotropic $s$-wave model has the value of $\Delta_1/\Delta_0$ smaller than $15\%$
which is negligible within the fitting precision. If the second model is accepted, the
simulations are also not bad as a whole, but can not fit very well the spectral shape
around zero bias.}
\end{figure}

In this section, we try to explain our experimental data with other possible models. In
Ref.\cite{KashiwayaS1998}, the authors ascribed the unphysically large $A$
($A=\Gamma/\Delta\sim 1$) to the unreasonable assumption of isotropic $s$-wave symmetry.
They found that if the maximum value of $\Gamma/\Delta$ was assumed to be 0.2 similar to
YBCO, reasonable fitting is obtained by assuming anisotropic $s$-wave symmetry, namely,
$\Delta(\theta)=\Delta_0+\Delta_1cos(4\theta)$. Comparing this previous report with
present work, we find that the large value of $A$ is mainly due to the surface
degradation. Nonetheless, the anisotropic $s$-wave symmetry seems to be reasonable
considering the crystallographic symmetry and the topology of Fermi surface (FS) of NCCO
\cite{KingDM1993,YuanQS2004}. In Fig.~\ref{fig:fig16}(a), the best fitting to such
anisotropic $s$-wave model is presented for both (110) and (100) directions. Recently,
the ARPES experiment revealed another form of the gap function described as
$\Delta(\theta)=\Delta_0\cdot|cos(2\theta)-0.3cos(6\theta)|$. Such $\Delta(\theta)$
function has also been tried in our fittings as shown in Fig.~\ref{fig:fig16}(b). All the
parameters for the theoretical simulations of Fig.~\ref{fig:fig16} are listed in Table
~\ref{tab:table2}. It is found that the isotropic $s$-wave is the best candidate among
the three $s$-wave models mentioned above.

\begin{table}
\caption{\label{tab:table2} List of the parameters for the fitting presented in
Fig.~\ref{fig:fig16}.}
\begin{ruledtabular}
\begin{tabular}{lccccc}
Label & $\Delta_0$ (meV) & $\Delta_1$ (meV) & $Z$ & $\Gamma$ (meV) & $\Gamma/\Delta$ \\
\hline
(a)-(110)  & 3.58 & 0.46 & 2.95 & 0.68 & 0.17  \\
(a)-(100)  & 3.45 & 0.44 & 2.80 & 0.84 & 0.21  \\
\hline \hline
%another& & & \\
\hline
(b)-(110)  & 4.50 & None & 3.20 & 0.54 & 0.12  \\
(b)-(100)  & 4.30 & None & 3.10 & 0.64 & 0.15  \\
\end{tabular}
\end{ruledtabular}
\end{table}

\begin{figure}[]
\includegraphics[scale=1.2]{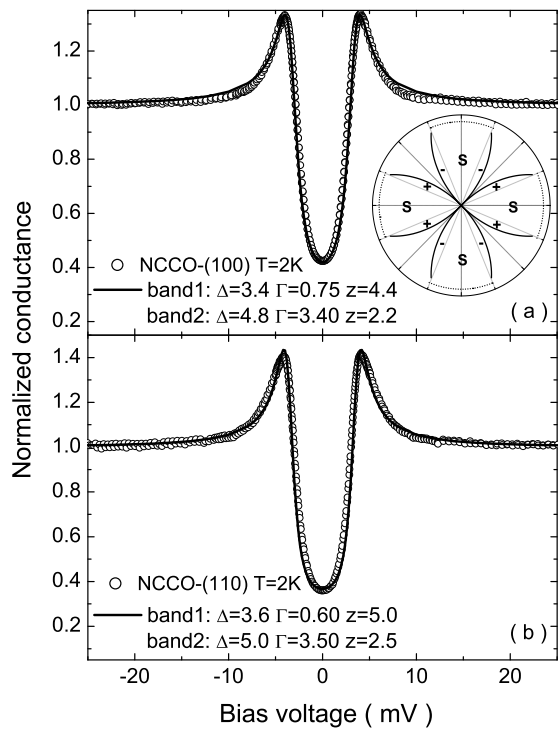}
\caption{\label{fig:fig17} Fitting the directional tunneling spectra to the two-band
model. (a) Along (100) direction; (b) Along (110) direction. All the fitting parameters
are listed in the corresponding panels. The angle dependent gap amplitude and sign are
described in the inset. Dotted lines indicate the $s$-wave component while the solid
lines indicate the $d$-wave component.}
\end{figure}

Recent ARPES experiments \cite{ArmitageNP2002,DamascelliA2003} revealed the doping
evolution of the Fermi surface (FS) in Nd$_{2-x}$Ce$_x$CuO$_4$. At low doping, a small
Fermi pocket appears around ($\pi$,0). Upon increasing doping, another pocket begins to
form around ($\pi/2$,$\pi/2$) and eventually at optimal doping $x=0.15$ several FS pieces
evolve into a large curve around ($\pi$,$\pi$). These findings were effectively described
as a two-band system \cite{YuanQS2004}. Most recently, Luo {\it et al.} used a weakly
coupled two-band BCS-like model to account for the low energy electromagnetic response of
superconducting quasiparticles in electron-doped materials \cite{LuoHG2004}. The special
angle dependence of the energy gap observed in Ref.\cite{MatsuiH2004} (as shown in the
inset of Fig.~\ref{fig:fig16}(b)) may be a reflection of the two-band model. In the
analysis in Ref.\cite{LuoHG2004}, both the pairing symmetries of band-1 and band-2 are
assumed to be the $d_{x^2-y^2}$ type in order to achieve the best fitting for the
superfluid density data, in which the labels 1 and 2 represent the bands contributing to
the FS centered at ($\pm\pi$,0), (0,$\pm\pi$) and ($\pm\pi/2$,$\pm\pi/2$), respectively.
However, as pointed out by the authors, there is a finite excitation gap in band-1 since
the nodal lines do not intersect with the FS of that band if the system is not heavily
overdoped. This indicates that the superconducting state of NCCO is actually a mixture of
$d$-wave and $s$-wave-like pairing states. In other words, this analysis can not
definitely determine whether the pairing symmetry of band-1 is $s$-wave or $d$-wave type.
Based on this two-band model, we calculated the directional tunneling spectra by assuming
different pairing symmetries ($s$-wave or $d$-wave) for band-1 while definite
$d_{x^2-y^2}$ for band-2. In these calculations, the ratio of the contribution from
band-1 to that from band-2 was chosen according to the discussions in
Ref.\cite{LuoHG2004}. If only the $d$-wave symmetry of band-1 is accepted, a prominent
ZBCP will appear in the spectra along (110) direction. That is, the $s$-wave symmetry of
band-1 is required in the fitting procedure. Moreover, the best fitting needs a much
larger value of $A=\Gamma/\Delta$ for band-2 than that for band-1 as shown in
Fig.~\ref{fig:fig17}, possibly due in part to the strong depression of the $d$-wave
superconductivity on the sample surface while the essential origin is yet to be found.
Such mixture of superconductivity coming from two bands may be another possible reason of
the contradicting reports on the pairing symmetry by different experiments. For example
the phase sensitive measurements \cite{TsueiCC2000PRL,Chesca} may selectively detect the
gap information from band-2 which crosses the Fermi surface near ($\pi/2,\pi/2$). In
addition, such inter-band mixture may also be responsible to the doping dependent pairing
symmetry observed in $Pr_{2-x}Ce_xCuO_4$ thin films \cite{BiswasA2002}. In any case, the
$s$-wave pairing symmetry appears to be an important component in the superconductivity
of optimally-doped NCCO.

In summary, the directional tunnelling spectra along (110) and (100) axis on the NCCO and
LSCO single crystals illustrate clearly distinct pairing symmetries. In contrast to the
results of LSCO (x=0.11), no ZBCP was observed for NCCO (x=0.15) along the two different
directions while sharp coherence peaks are existent for both directions, which disagrees
with the pure $d$-wave pairing symmetry. The almost identical spectral shapes for the two
directions on NCCO can be understood in the framework of $s$-wave BTK theory, leading to
a BCS type temperature dependence of energy gap with the ratio of $\Delta/k_B T_c\approx
1.66$. The present work provides evidence for the $s$-wave component in the
superconductivity of the optimally-doped NCCO, which should mainly come from the band
crossing the FS around ($\pm\pi$,0) and (0,$\pm\pi$) if a two-band model is accepted.

\begin{acknowledgments}
The authors thank Professor T.Xiang and Dr. H.G.Luo for fruitful discussions. The work is
supported by the National Science Foundation of China, the Ministry of Science and
Technology of China, and Chinese Academy of Sciences with the Knowledge Innovation
Project. Pengcheng Dai is supported by the U. S. NSF DMR-0139882, DOE No.
DE-AC05-00OR22725 with UT/Battelle, LLC., and by NSF of China under contract No.
10128408. We thank Dr. Y.G.Shi for the provision of some chemicals.
\end{acknowledgments}

E-mail address: hhwen@aphy.iphy.ac.cn (H. H. Wen); shanlei@ssc.iphy.ac.cn (L. Shan)

%\bibliography{NoEndingPoint}

\end{document}